\documentclass[pra,aps,twocolumn,nopacs,superscriptaddress,nofootinbib]{revtex4}
\usepackage[bbgreekl]{mathbbol}
\usepackage{appendix}
\usepackage[T1]{fontenc}
\usepackage[latin1]{inputenc}
\usepackage{lmodern}
\usepackage{graphicx}
\usepackage{dcolumn}
\usepackage{bm}
\usepackage{amsmath}
\usepackage{amssymb}
\usepackage{pst-all}
\usepackage{psfrag}
\usepackage{epsfig}
\usepackage{titletoc}
\usepackage{minitoc}

\def\ii{{\rm i}} \def\ee{{\rm e}}  
\def\Eb{{\bf E}} \def\Rb{{\bf R}} \def\rb{{\bf r}}  
  
 \def\pb{{\bf p}}

  \def\eh{{\hat{\bf e}}}

\begin{document}

\title{Plasmon-enhanced nonlinear wave mixing in nanostructured graphene}

\author{Joel D. Cox}
\email{joel.cox@icfo.es}
\affiliation{ICFO-Institut de Ci\'encies Fot\'oniques, Mediterranean Technology Park, 08860 Castelldefels (Barcelona), Spain}
\author{F. Javier Garc\'{\i}a de Abajo}
\email{javier.garciadeabajo@icfo.es}
\affiliation{ICFO-Institut de Ci\'encies Fot\'oniques, Mediterranean Technology Park, 08860 Castelldefels (Barcelona), Spain}
\affiliation{ICREA-Instituci\'o Catalana de Recerca i Estudis Avan\c{c}ats, Passeig Llu\'{\i}s Companys 23, 08010 Barcelona, Spain}

\begin{abstract}
Localized plasmons in metallic nanostructures have been widely used to enhance nonlinear optical effects due to their ability to concentrate and enhance light down to extreme-subwavelength scales. As alternatives to noble metal nanoparticles, graphene nanostructures can host long-lived plasmons that efficiently couple to light and are actively tunable via electrical doping. Here we show that doped graphene nanoislands present unique opportunities for enhancing nonlinear optical wave-mixing processes between two externally applied optical fields at the nanoscale. These small islands can support pronounced plasmons at multiple frequencies, resulting in extraordinarily high wave-mixing susceptibilities when one or more of the input or output frequencies coincide with a plasmon resonance. By varying the doping charge density in a nanoisland with a fixed geometry, enhanced wave mixing can be realized over a wide spectral range in the visible and near infrared. We concentrate in particular on second- and third-order processes, including sum and difference frequency generation, as well as on four-wave mixing. Our calculations for armchair graphene triangles composed of up to several hundred carbon atoms display large wave mixing polarizabilities compared with metal nanoparticles of similar lateral size, thus supporting nanographene as an excellent material for tunable nonlinear optical nanodevices.
\end{abstract}
\maketitle
\tableofcontents


\section{Introduction}

\begin{figure*}[t!]
\includegraphics[width=0.9\textwidth]{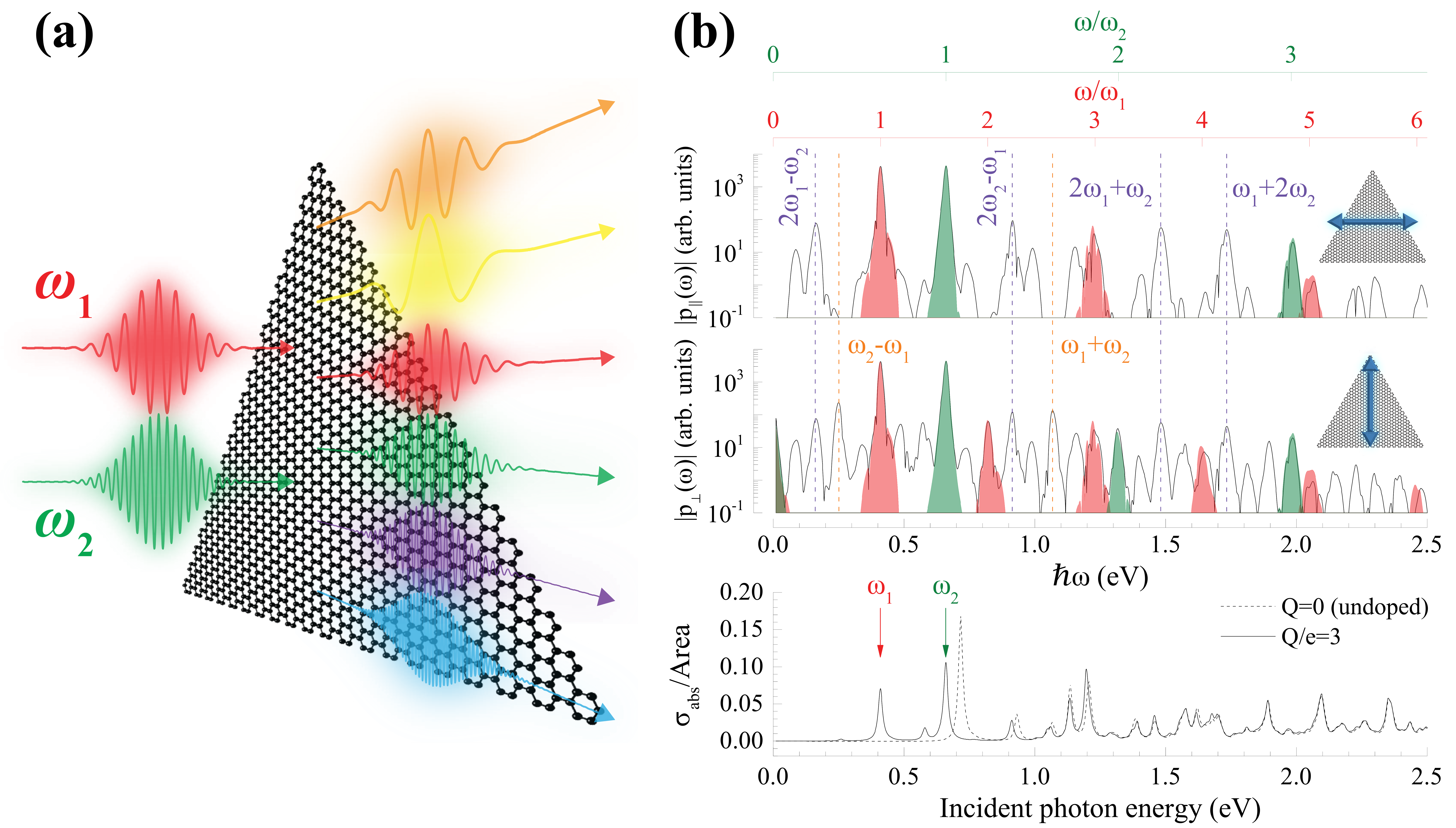}
\caption{\label{fig1} \textbf{Plasmon-enhanced wave mixing of coincident pulses.} \textbf{(a)} Illustration of an armchair triangular graphene nanoisland containing $N=1260$ carbon atoms (side length of $8.4\,$nm) and doped with charge $Q=3e$, illuminated by two collinear light pulses of central energies $\hbar\omega_1=0.41\,$eV and $\hbar\omega_2=0.66\,$eV, each having $200\,$fs FWHM duration and peak intensity $10^{12}\,$W/m$^2$. \textbf{(b)} Spectral density of the induced dipole moment under excitation by light polarized along a direction parallel (upper panel) or perpendicular (middle panel) to one of the nanotriangle sides (dipole along the incident electric-field direction). The solid black curves correspond to the spectra obtained upon excitation by coincident pulses, while the filled curves show the spectra produced by individual pulses of central frequency $\omega_1$ (red) or $\omega_2$ (green). The linear absorption cross-section of the nanoisland (normalized to its area) is presented in the lower panel (solid curve), where the arrows indicate the central frequencies of the exciting pulses and we compare it with the absorption of the undoped island (dashed curve).}
\end{figure*}

\begin{figure*}[t]
\includegraphics[width=1\textwidth]{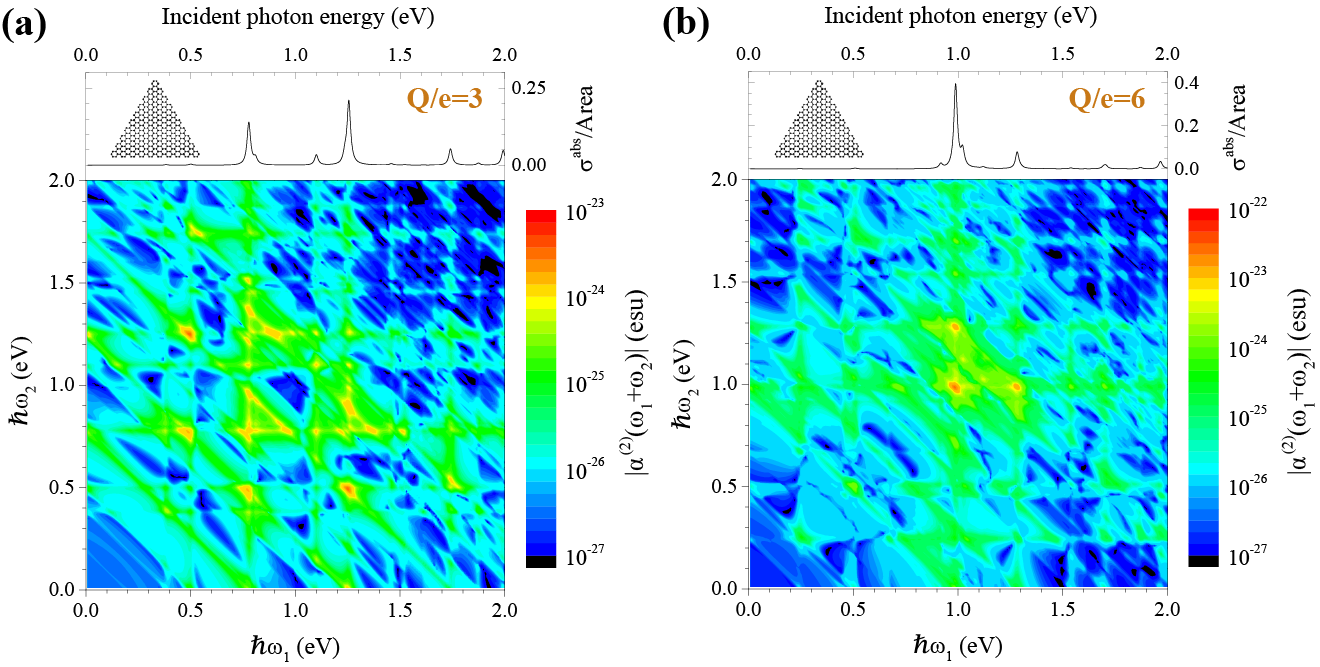}
\caption{\label{fig2} \textbf{Sum frequency generation} We show the nonlinear polarizability $\alpha^{(2)}(\omega_1+\omega_2)$, corresponding to sum-frequency generation (SFG), for an armchair-edged triangular graphene nanoisland containing $N=330$ carbon atoms and doped with (a) three and (b) six additional electrons. The upper plots show the linear absorption cross-section, while the lower panels show the polarizabilities as the incident field frequencies $\omega_1$ and $\omega_2$ are varied.}
\end{figure*}

Nonlinear optical phenomena arise from the coupling between two or more photons, mediated by their interaction with matter, to produce a photon with a frequency that is a linear combination of the original photon frequencies. These processes are responsible for many significant advances in laser-based technologies \cite{B08_3,G13}, most of which rely on phase-matching of intense electromagnetic fields in extended bulk crystalline media to enable efficient frequency conversion. Now, as the accessibility of nanostructured materials offered by modern nanofabrication techniques continues to increase, so does the interest in mastering nonlinear optics on subwavelength scales. Indeed, nonlinear optical nanomaterials find diverse applications, including optical microscopy \cite{WLN11,HC13}, biological imaging/detection \cite{PRH10,PMW10}, and signal conversion in nanoscale photonic devices \cite{NV11,CAA12,ARN14,MHS14}.

Inherently, nanostructured materials possess small volumes that limit their interaction with optical fields. Fortunately, this can be compensated by high oscillator strengths provided by either quantum confinement effects \cite{A96} or the intense near-fields generated by localized surface plasmons \cite{NV11,S11}. In particular, the near-field enhancement associated with plasmons in noble metal nanostructures has been widely used with the purpose of enhancing the nonlinear response of surrounding dielectric materials \cite{KZ12,ARN14,MHS14}. Plasmons also display strong intrinsic optical nonlinearities, clearly observed in metal nanoparticles \cite{DN07,PDN09,KZ12,V12}.

The plasmonic response of a metal nanostructure can be tailored by its size, shape, and surrounding environment \cite{S11}, as well as by combining two or more interacting nanostructures. In this manner, a metallic nanocomposite can be engineered to possess multiple resonance frequencies, enabling the optimization of nonlinear wave mixing between multifrequency optical fields \cite{HVQ12,ZWZ13}. Unfortunately, plasmons are hardly tunable in metals after a structure has been fabricated \cite{NH06}, thus severely limiting the choice of frequency combinations.

As an alternative plasmonic material, electrically doped graphene has been found to support long-lived plasmonic excitations that efficiently couple to light and are actively tunable by changing the density of charge carriers \cite{JGH11,FAB11,paper196,FRA12,YLC12,YLL12,paper212,BJS13,paper230,YLZ13,FLZ14,GPN12,paper235}. Strong intrinsic nonlinearities have also been observed in this material \cite{HHM10,ZVB12,KKG13}, which could be further enhanced by plasmons \cite{M11,paper226}. Recently, plasmon-assisted second-harmonic generation (SHG) and down conversion with good efficiencies at the few-photon level have been shown to be possible when the fundamental and the second-harmonic are simultaneously resonant with plasmons in a graphene nanoisland \cite{MSG14}. We have predicted that a similar mechanism can lead to unprecedentedly intense SHG and third-harmonic generation (THG) in nanographene \cite{paper247}. Although experimental studies have so far demonstrated strong plasmons at mid-infrared and terahertz frequencies \cite{JGH11,FAB11,paper196,FRA12,YLC12,YLL12,paper212,BJS13,paper230,YLZ13,FLZ14}, further reduction in the size of the islands down to less than 10\,nm should allow us to reach the visible and near-infrared (vis-NIR) regimes \cite{paper183,paper214,paper215,paper235}. In particular, commercially-available polycyclic aromatic molecules sustain plasmon-like resonances that are switched on and off by changing their charge state \cite{paper215}. Additionally, recent progress in the chemical synthesis of nanographene \cite{WPM07,FLP08,FPM09,CRJ10,DHZ14} provides further stimulus for the use of this material to produce electrically tunable vis-NIR plasmons, as well as their application to nonlinear optics at the nanoscale.

In this work, we investigate nonlinear optical wave mixing in doped nanographene. Specifically, we propose a scheme for optimizing wave mixing among multifrequency optical fields that utilizes the plasmons supported by a doped graphene nanoisland. Using a tight-binding description for the electronic structure combined with density-matrix quantum-mechanical simulations, we demonstrate that efficient wave mixing is achieved in an island when the incident and/or the mixed frequencies are coupled with one or more of its plasmons. By actively tuning the doping level in a graphene nanoisland of fixed geometry, a wide range of plasmon-enhanced input/output mixing frequency combinations can be realized.

\section{Results and discussion}

We study wave mixing of coincident light pulses in Fig.\ \ref{fig1} for a triangular armchair-edged graphene nanoisland containing $N=1260$ carbon atoms (side length pf $8.4\,$nm) and doped with three electrons (doping density $9.1\times10^{12}\,\text{cm}^{-2}$, corresponding to an equivalent extended graphene Fermi energy $E_F=0.35\,$eV). Additionally, we assume a conservative inelastic lifetime $\tau=33\,$fs (i.e., linewidth $\hbar\tau^{-1}=20\,$meV). From the linear absorption cross-section of the nanoisland, we identify several prominent plasmons for incident photon energies. In particular, we concentrate on $\hbar\omega_1=0.41\,$eV and $\hbar\omega_2=0.66\,$eV. The first of these plasmons is highly tunable upon electrical doping, as it is switched on when moving from $Q=0$ to $Q\neq0$ charge states (see lower panel in Fig.\ \ref{fig1}b), and its frequency increases with $Q$ (see Fig.\ \ref{figS1} in the Appendix). The plasmon at $\omega_2$ is comparatively less tunable. Plasmon-enhanced optical wave mixing in the nanoisland is demonstrated in Fig.\ \ref{fig1}b, where the spectral decomposition of the induced dipole moment is shown for excitation by collinear Gaussian pulses with central frequencies $\omega_1$ and $\omega_2$. We show the response for incident polarization aligned with an edge of the nanoisland (upper panels) or perpendicular to an edge (middle panels), and in each case the dipole is calculated along the direction of polarization. Further, in Fig.\ \ref{fig1}b we superimpose the spectra obtained from coincident pulses with the spectra for excitation of the nanoisland by each of the individual pulses in isolation, which exhibit polarization features that oscillate at harmonics of the fundamental frequencies ($n\omega_1$ and $n\omega_2$). The dual-pulse spectrum exhibits these features in addition to polarization components produced via sum and difference frequency generation ($\omega_1\pm\omega_2$) and degenerate four-wave mixing ($\omega_1\pm2\omega_2$ or $2\omega_1\pm\omega_2$), along with various other combinations of harmonic generation and wave mixing. We note that for polarization along the edge of a nanotriangle, inversion symmetry in this direction prevents even-ordered nonlinear processes from occurring, such as SHG and sum/difference frequency generation. Incidentally, mixing of less tunable plasmons (e.g., $\hbar\omega_2$ with the resonance at $1.20\,$eV) produces less intense nonlinear features (see Fig.\ \ref{figS2} in the Appendix).

\begin{figure*}[t]
\includegraphics[width=1\textwidth]{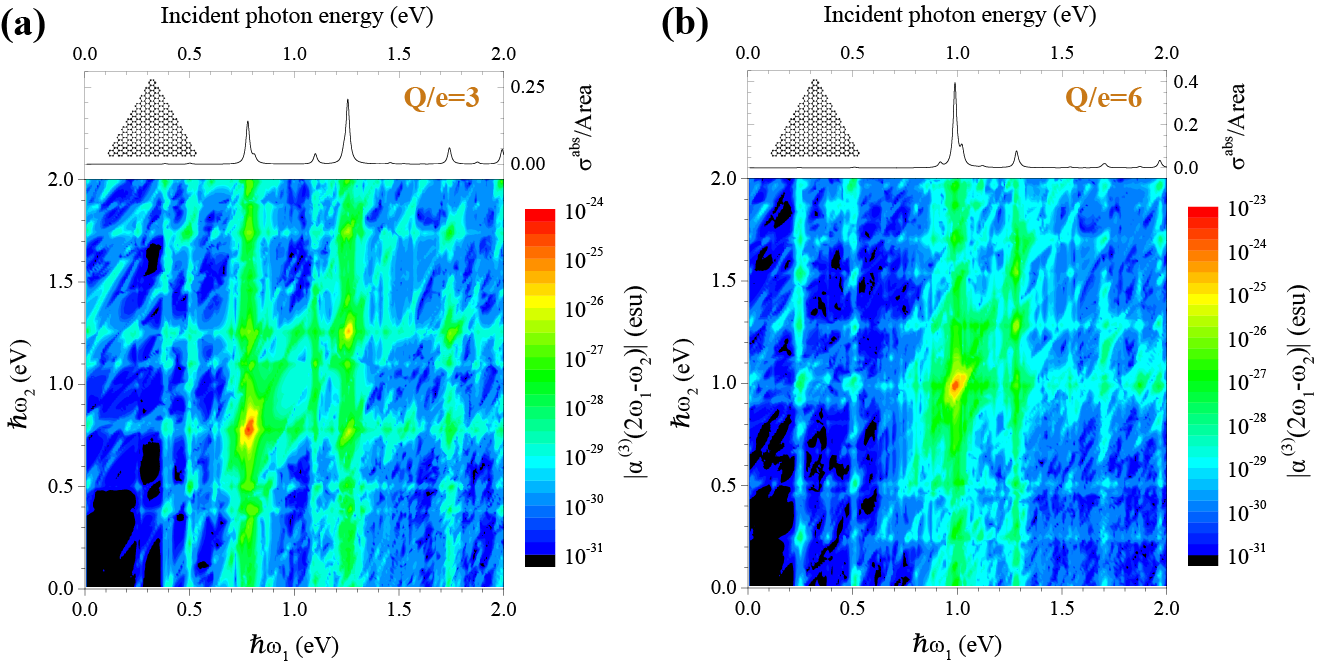}
\caption{\label{fig3} \textbf{Four-wave mixing} We show the nonlinear polarizability $\alpha^{(3)}(2\omega_1-\omega_2)$, corresponding to four-wave mixing, for the same nanoisland and doping conditions considered in Fig.\ \ref{fig2}.}
\end{figure*}

While the response of a graphene nanoisland to ultrashort pulses can provide information on the relative strengths of the nonlinear processes for excitation at specific frequencies, a quantitative analysis of each nonlinear process, along with its optimal plasmonic enhancement, is best provided by studying the response under continuous-wave (cw) illumination. In Fig.\ \ref{fig2} we consider sum-frequency generation (SFG) in a nanoisland containing $N=330$ atoms (side length of $4.4\,$nm), for doping with either three (Fig.\ \ref{fig2}a) or six (Fig.\ \ref{fig2}b) additional charge carriers. The upper panels in Fig.\ \ref{fig2} show the linear response of the nanoisland for the two doping levels considered, which show multiple plasmon resonance peaks at low doping that converge to a single, stronger feature as the doping level increases. The SFG polarizability $\alpha^{(2)}(\omega_1+\omega_2)$ is presented as a function of the two applied field frequencies, enabling exploration of all possible frequency combinations that may result in plasmon-enhanced wave mixing.

The input frequencies at which SFG is enhanced are found to coincide with the plasmons, corresponding to the horizontally and vertically aligned features in the contour plot of $|\alpha^{(2)}(\omega_1+\omega_2)|$ in Fig.\ \ref{fig2}. Strong enhancement is observed where these features intersect, driven by plasmons excited at both fundamental frequencies. Prominent features also follow frequencies satisfying $\omega_2+\omega_1=\omega_p$, where $\omega_p$ denotes one of the plasmon frequencies of the nanoisland, indicating plasmonic enhancement at the output (sum) frequency. Finally, since SHG can be considered as a special case of SFG, we also note plasmonic enhancement of SHG for frequencies satisfying $\omega_1=\omega_2=\omega_p$.

Similar conclusions are drawn from frequency-difference generation (i.e., $\alpha^{(2)}(\omega_1-\omega_2)$), where we observe enhancement when either one of the incident frequencies or the output frequency resonates with a plasmon (see Fig.\ \ref{figS3} in the Appendix).

Plasmonic enhancement of four-wave mixing is investigated in Fig.\ \ref{fig3} for the same graphene nanoisland and doping conditions considered for Fig.\ \ref{fig2}, with an output frequency $2\omega_1-\omega_2$. We find a similar enhancement in $\alpha^{(3)}(2\omega_1-\omega_2)$ when either of the fundamental frequencies of the incident fields are resonant with plasmons in the nanoisland, although in this case the enhancement favors $\omega_1=\omega_p$, as $\omega_1$ is involved twice in these specific wave mixing processes. Here we find increased four-wave mixing following the frequencies $\omega_2=2\omega_1-\omega_p$, once again indicating enhancement at the output frequency. For $\omega_1=\omega_2$ we have fully-degenerate four-wave mixing, where we are actually investigating the third-order polarizability $\alpha^{(3)}(\omega_1)$ contributing to the linear response (i.e., the Kerr effect). In this case, frequencies satisfying the condition $\omega_1=\omega_2=\omega_p$ produce extremely high polarizabilities, as the three input frequencies and the output frequency are all equally amplified by the same plasmon resonance. We have also examined four-wave mixing with output frequency $2\omega_1+\omega_2$, leading to similar conclusions on plasmon-assisted enhancement (see Fig.\ \ref{figS4} in the Appendix).

\begin{figure*}[t!]
\includegraphics[width=1\textwidth]{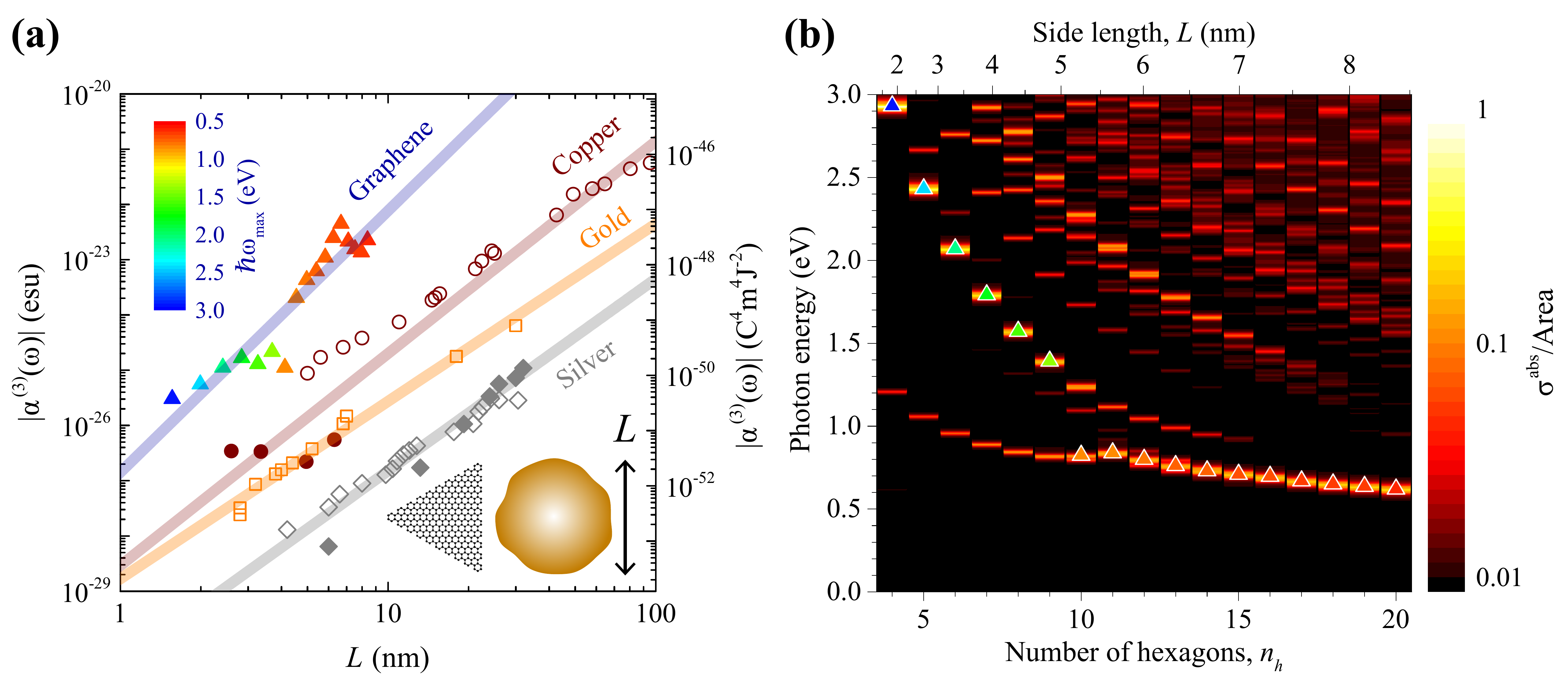}
\caption{\label{fig4} \textbf{Third-order response of nanographene compared to that of noble metal nanoparticles.} (a) The maximum nonlinear polarizabilities $\alpha^{(3)}(\omega)$ of graphene nanoislands (solid triangles) as a function of their side length $L$ are compared with measured values of noble metal nanoparticles of similar diameter reported in the literature. The color scale indicates the incident photon energy at which the maximum graphene nonlinear polarizabilities are found. The semi-transparent solid lines are guides to the eye, added to indicate the scaling with $L$. (b) Linear absorption spectra for the armchair-edged graphene nanotriangles considered in (a), distinguished by the number of hexagons along the edge $n_h$, for which a fixed doping density of one electron per 100 carbon atoms is maintained. The frequencies at which the maximum values of $\alpha^{(3)}(\omega)$ occur are indicated by the solid triangles, which follow the color scale in (a). The experimental data used in (a) is extracted from Refs.\ \cite{HRF1988,UKO94,YBS94,SOO14}, where the polarizabilities are obtained upon multiplication of the reported third-order susceptibility $|\chi^{(3)}|$ by the particle volume. Specifically, we show data obtained via fully degenerate four-wave mixing measurements of several noble metals performed at different wavelengths: gold nanoparticles at $\sim530\,$nm \cite{HRF1988} (hollow squares), silver nanoparticles at $\sim420\,$nm \cite{UKO94} (hollow diamonds) and near $\sim410\,$nm \cite{SOO14} (solid diamonds), and copper nanoparticles at $\sim570\,$nm \cite{UKO94} (hollow circles) and $\sim530\,$nm \cite{YBS94} (solid circles).}
\end{figure*}

As noble metal nanoparticles possess the highest nonlinear polarizabilities per atom measured experimentally \cite{KZ12}, we compare them with the above results for doped nanographene nonlinear polarizabilities. In particular, we contrast data available in the literature for metal third-order polarizabilities $\alpha^{(3)}(\omega)$ with those calculated here for graphene. In Fig.\ \ref{fig4}a we show the maximum polarizability $|\alpha^{(3)}(\omega)|$ for graphene nanoislands of increasing size, and for a fixed doping ratio of one electron per every hundred carbon atoms, along with experimental data obtained from degenerate four-wave mixing experiments on gold, silver, and copper nanoparticles of comparable or greater sizes. Even through metal particles have much larger volumes than graphene nanoislands for a given lateral size $L$, we find the latter to exhibit much larger nonlinear response, reaching two orders of magnitude higher values. Obviously, we are comparing theory for graphene with experiments for metals, so this is a preliminary conclusion, which should be examined experimentally. Nonetheless, we are using a conservative inelastic lifetime for nanographene in our calculations ($\tau=33\,$fs), whereas much longer lifetimes could be actually encountered in practice, based upon recently measured plasmon resonances \cite{paper212,GRARPATD2}, and it should be noted that the nonlinear polarizability of order $n$ scales roughly as $\tau^n$ \cite{paper247}. The linear absorption spectra for the nanoislands that we compare with metal nanoparticles are presented in Fig.\ \ref{fig4}b, illustrating how the plasmon resonances vary as the nanoisland size increases while a fixed doping density is maintained.

\section{Conclusions}

In conclusion, graphene nanoislands can be used to realize nonlinear wave mixing on the nanoscale with extraordinarily high efficiencies, surpassing those of metal nanoparticles of similar lateral sizes. The large magnitudes of the predicted nonlinear polarizabilities are attributed to plasmonic enhancement in these nanostructures. Wave mixing is further enhanced by simultaneously exploiting multiple plasmonic resonances, enabling two incident fields with distinct frequencies to independently couple to plasmons, and additionally tuning the mixed frequency to yet another plasmon. These plasmons can be tuned by changing the number of doping charge carriers, adding another advantage with respect to conventional plasmonic metal response. Interestingly, by considering islands formed by a few hundreds or thousands of carbon atoms, we predict tunable plasmonic response and plasmon-induced enhanced nonlinearities within the visible and near-infrared spectral ranges. Our results configure a new platform for the development of nanoscale nonlinear optical devices based upon graphene nanostructures with lateral dimensions of only a few nanometers, such as those that are currently produced by chemical synthesis \cite{WPM07,FLP08,FPM09,CRJ10,DHZ14}.

\section{Methods}

We describe the low-energy ($<3\,$eV) optical response of graphene nanoislands within a density-matrix approach, using a tight-binding model for the $\pi$-band electronic structure \cite{W1947,CGP09}. One-electron states $|\varphi_j\rangle$ are obtained by assuming a single $p$ orbital per carbon site, oriented perpendicular to the graphene plane, with a hopping energy of $2.8\,$eV between nearest neighbors. In the spirit of the mean-field approximation \cite{HL1970}, a single-particle density matrix is constructed as $\rho=\sum_{jj'}\tilde{\rho}_{jj'}|\varphi_j\rangle\langle\varphi_{j'}|$, where $\tilde{\rho}_{jj'}$ are time-dependent complex numbers. An incoherent Fermi-Dirac distribution of occupation fractions $f_j$ is assumed in the unperturbed state, characterized by a density matrix $\tilde{\rho}^0_{jj'}=\delta_{jj'}f_j$, whereas the time evolution under external illumination is governed by the equation of motion
\begin{equation}\label{rho_eom}
\frac{\partial \rho}{\partial t} = -\frac{i}{\hbar} \left[ H, \rho \right] - \frac{1}{2\tau} \left( \rho - \rho^{0} \right).
\end{equation}
The last term of Eq.\ \ref{rho_eom} describes inelastic losses at a phenomenological decay rate $1/\tau$. We set $\hbar\tau^{-1}=20\,$meV throughout this work (i.e., $\tau=33\,$fs), corresponding to a conservative Drude-model graphene mobility $\mu\approx460\,$cm$^2\,$V$^{-1}$\,s$^{-1}$ for a characteristic doping carrier density $4\times10^{13}\,$cm$^{-2}$ (i.e., one charge carrier per every 100 carbon atoms). The system Hamiltonian $H=H_{\rm TB}-e\phi$ consists of the tight-binding part $H_{\rm TB}$ (i.e., nearest-neighbors hopping) and the interaction with the self-consistent electric potential $\phi$, which is in turn the sum of external and induced potentials. The latter is simply taken as the Hartree potential produced by the perturbed electron density, while the former reduces to $-\rb\cdot\Eb(t)$ for an incident electric field $\Eb(t)$. The induced dipole moment is then calculated from the diagonal elements of the density matrix in the carbon-site representation as $\pb(t)=-2e\sum_l\big[\rho_{ll}(t)-\rho^0_{ll}\big]\Rb_l$, where the factor of 2 accounts for spin degeneracy and $\Rb_l=(x_l,y_l)$ runs over carbon sites.

We use two different methods to solve Eq.\ \ref{rho_eom} and find $\pb(t)$ (direct time-domain numerical integration and a perturbative approach \cite{paper247}), which we find in excellent mutual agreement under low-intensity cw illumination. Direct time integration allows us to simulate the response to short light pulses for arbitrarily large intensity, while the perturbative method yields the nonlinear polarizabilities under multifrequency cw illumination ($\Eb(t)=E_0\,\hat{\bf e}\,\ee^{-\ii\omega_1 t}+E_0\,\hat{\bf e}\,\ee^{-\ii\omega_2 t}+{\rm c.c.}$, with the same amplitude $E_0$ at both frequencies for simplicity), for which we can express the dipole moment as a power series in the electric field strength $E_0$ according to
\begin{align}
\pb(t)=&\sum_{n}\sum^{n}_{\substack{s_1=-n\\s_2=-n}} \alpha^{(n)}(s_1\omega_1+s_2\omega_2)\,(E_0)^n\,\ee^{-\ii (s_1\omega_1+s_2\omega_2) t}\nonumber \\ &+{\rm c.c.}\label{pt}
\end{align}
Here, $n$ is the scattering order, whereas $s_1$ and $s_2$ give the harmonic orders of the incident frequencies $\omega_1$ and $\omega_2$. We denote polarizabilities according to $\alpha^{(n)}(\omega_{\rm out})$, where $\omega_{\rm out}$ is the frequency generated by a particular $n^{\rm th}$-order process. Considering terms up to third order ($n\le3$), Eq.\ \ref{pt} then defines the linear polarizability $\alpha^{(1)}(\omega_i)$ ($i=1$ or 2), the polarizabilities for SHG and THG, $\alpha^{(2)}(2\omega_i)$ and $\alpha^{(3)}(3\omega_i)$, respectively, the wave-mixing polarizabilities corresponding to sum and difference frequency generation $\alpha^{(2)}(\omega_i\pm\omega_j)$ ($j=1$ or 2, $i\ne j$), and the four-wave mixing polarizabilities $\alpha^{(3)}(2\omega_i\pm\omega_j)$ and $\alpha^{(3)}(\omega_i+\omega_j-\omega_j)$. We obtain these polarizabilities by expanding the density matrix in Eq.\ \ref{rho_eom} as 
\begin{equation}\label{rhons1s2}
\rho=\sum_n\sum^{n}_{\substack{s_1=-n\\s_2=-n}}\rho^{ns_1s_2}\ee^{-\ii (s_1\omega_1+s_2\omega_2)t},
\end{equation}
which leads to self-consistent equations in $\rho^{ns_1s_2}$ (for the relevant combinations of $n$ with harmonics $s_1$ and $s_2$). We solve Eq.\ \ref{rhons1s2} following a procedure inspired in the random-phase approximation (RPA) formalism, as discussed elsewhere \cite{paper247} for SHG and THG. Further details on the extension of this formalism to cope with wave mixing are given in the Appendix.

\acknowledgments

This work has been supported in part by the European Commission (Graphene Flagship CNECT-ICT-604391 and FP7
-ICT-2013-613024-GRASP).



\appendix
\appendixpage
\addappheadtotoc

The perturbative method used to simulate the optical response of nanographene to multifrequency continuous wave illumination is presented in detail. We also include additional results from numerical simulations performed in the time-domain, and provide further wave-mixing polarizability spectra for graphene nanoislands with various sizes and dopings.

\section{Perturbative method applied to wave mixing}

\begin{figure*}
\includegraphics[width=0.9\textwidth]{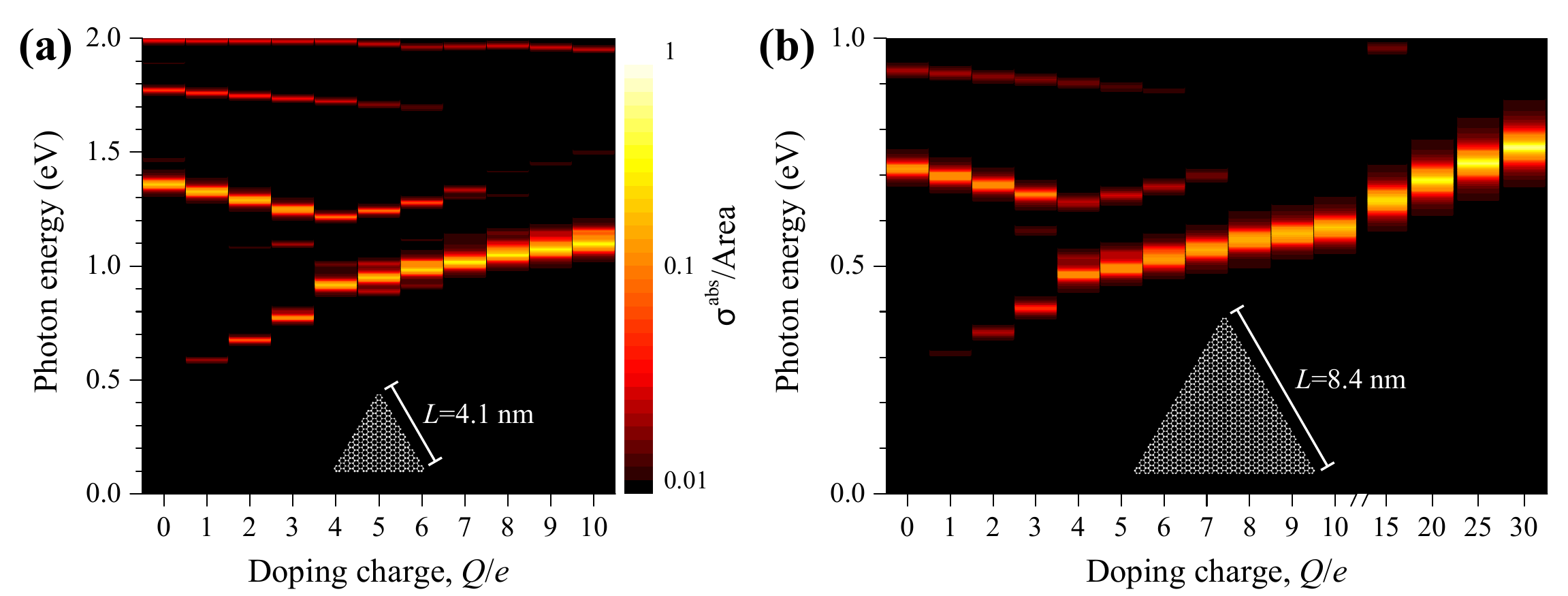}
\caption{\label{figS1} \textbf{Linear absorption spectra for armchair nanotriangles.} Spectra for triangular, armchair-edged graphene nanoislands with side lengths (a) $L=4.1\,$nm and (b) $L=8.4\,$nm are shown as a function of the number of additional charge carriers $Q/e$. The absorption spectra are normalized to the area of the triangle. We use a common color scale in (a) and (b).}
\end{figure*}

Following a procedure similar to that described in previous work for the analysis of multiple-harmonic generation \cite{paper247}, we express the single-particle density matrix equation of motion for a graphene nanoisland in the basis set of its tight-binding electronic states as
\begin{align}\label{dmjj}
\frac{\partial \tilde{\rho}_{jj'}}{\partial t} =& -\ii\left(\varepsilon_j - \varepsilon_{j'}\right)\tilde{\rho}_{jj'} \\
&+\frac{\ii e}{\hbar}\sum_{l,l'} \left( \phi_l - \phi_{l'} \right)a_{jl}a_{j'l'}\rho_{ll'} -\frac{1}{\tau}\left( \tilde{\rho}_{jj'}-\tilde{\rho}^{0}_{jj'}\right),
\nonumber
\end{align}
where $\hbar\varepsilon_j$ is the energy of state $|\varphi_j\rangle$ and $\phi_l=\langle l|\phi|l \rangle$ are the matrix elements of the total electric potential in the basis set of the 2p orbitals $|l\rangle$ located at the carbon sites $\textbf{R}_l$. The single-electron states and the carbon site orbitals are related through
\begin{equation}\label{jstate}
|\varphi_j\rangle=\sum_l a_{jl}|l\rangle,
\end{equation}
where the real-valued coefficients $a_{jl}$ give the amplitude of orbitals $|l\rangle$ in states $|\varphi_j\rangle$. These states are orthonormal ($\sum_l a_{jl}a_{j'l}=\delta_{jj'}$) and form a complete set ($\sum_j a_{jl}a_{jl'}=\delta_{ll'}$), thus facilitating transformations of the density matrix elements between site and state representations according to $\tilde{\rho}_{jj'}=\sum_{ll'}a_{jl}a_{j'l'}\rho_{ll'}$ and $\rho_{ll'}=\sum_{jj'}a_{jl}a_{j'l'}\tilde{\rho}_{jj'}$. In what follows, we use indices $l$ to label carbon sites and $j$ for single-electron states.

\begin{figure*}[t]
\includegraphics[width=1\textwidth]{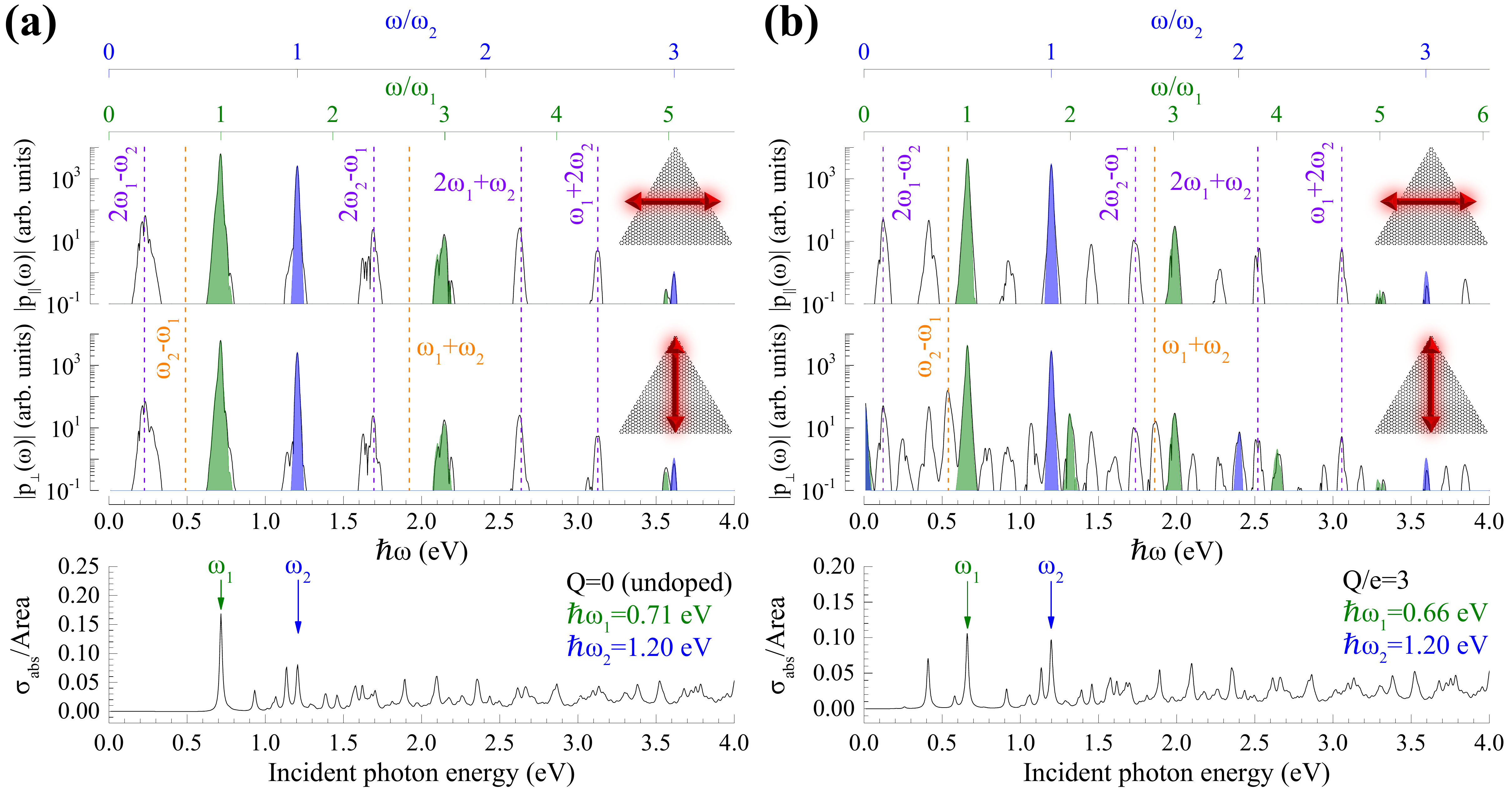}
\caption{\label{figS2} \textbf{Plasmon-enhanced wave mixing of coincident pulses.} We simulate the optical response of the nanoisland studied in Fig.\ 1 of the main text ($N=1260$ carbon atoms), for excitation by light pulses with the same duration ($200\,$fs FWHM) and peak intensity ($10^{12}\,$W/m$^2$). In (a) we show the induced dipole moment when the island is undoped and the incident pulses are tuned to the two most prominent resonances. In (b) we show the response for the same doping level considered in Fig.\ 1 of the main text (three additional electrons), but with $\omega_1$ and $\omega_2$ tuned to different resonances.}
\end{figure*}

In what follows, we solve Eq.\ (\ref{dmjj}) for graphene nanoislands exposed to multifrequency continuous-wave (cw) illumination described by the electric field
\begin{equation}\label{EE0}
\textbf{E}(t)=E_0\left(\ee^{-\ii\omega_1 t} + \ee^{-\ii\omega_2 t} + \text{c.c.}\right)\,\eh,
\end{equation}
where $E_0$ is the field amplitude and $\textbf{e}$ the polarization unit vector. Assuming that $E_0$ is weak, we expand the density matrix as
\begin{equation}\label{rhons}
\rho=\sum_{n,s_1,s_2}\rho^{ns_1s_2}\ee^{-\ii\left(s_1\omega_1+s_2\omega_2\right)t},
\end{equation}
where $n=1,2,3,...$ indicates the perturbation order (i.e., terms proportional to $(E_0)^n$, see Eq.\ (\ref{EE0})) and $s_1$ ($s_2$) is the harmonic index of $\omega_1$ ($\omega_2$). At $0^{\text{th}}$ order, Eq.\ (\ref{dmjj}) is trivially satisfied with $\rho^{0s_1s_2}=\delta_{s_1,0}\delta_{s_2,0}\rho^{0}$. Following Ref.\ \cite{paper247}, we insert Eq.\ (\ref{rhons}) into Eq.\ (\ref{dmjj}) and collect terms with the same perturbation order and $\ee^{-\ii\left(s_1\omega_1+s_2\omega_2\right)t}$ dependence, from which we obtain the density matrix at order $n\ge1$ as
\begin{equation}\label{rhons2}
\tilde{\rho}^{ns_1s_2}_{jj'}=-\frac{e}{\hbar}\sum_{l,l'}\frac{\left(\phi^{ns_1s_2}_l-\phi^{ns_1s_2}_{l'}\right)a_{jl}a_{j'l'}\rho^{0}_{ll'}}
{s_1\omega_1+s_2\omega_2+\ii/2\tau-\left(\varepsilon_j-\varepsilon_{j'}\right)}+\eta^{ns_1s_2}_{jj'},
\end{equation}
where
\begin{widetext}
\begin{equation}\label{etajj}
\eta^{ns_1s_2}_{jj'}=-\frac{e}{\hbar}\sum^{n-1}_{n'=1}\sum^{n'}_{\substack{s_1'=-n'\\s_2'=-n'}}\sum_{l,l'}
\frac{\left(\phi^{n's_1's_2'}_l-\phi^{n's_1's_2'}_{l'}\right)a_{jl}a_{j'l'}}
{s_1\omega_1+s_2\omega_2+\ii/2\tau-\left(\varepsilon_j-\varepsilon_{j'}\right)}\rho^{n-n',s_1-s_1',s_2-s_2'}_{ll'},
\end{equation}
\end{widetext}
and
\begin{align}\label{phinsrho}
\phi^{ns_1s_2}_{l}=&-(\textbf{R}_l\cdot\eh)\;E_0\delta_{n,1}\sum_{i=1}^{2}(\delta_{s_i,-1}+\delta_{s_i,1}) \\
&-2e\sum_{l'} v_{ll'} \rho^{ns_1s_2}_{l'l'} \nonumber
\end{align}
is the total potential. In Eq.\ (\ref{phinsrho}), the first term represents the contribution from the external field (nonzero only for order $n=1$), while the second term describes the Hartree potential produced by the perturbed electron density (here $v_{ll'}$ indicates the spatial dependence of the Coulomb interaction between electrons in orbitals $|l\rangle$ and $|l'\rangle$). The latter quantity ensures a linear dependence on $\rho^{ns_1s_2}$ for the first term on the right-hand side of Eq.\ (\ref{rhons2}), whereas lower perturbation orders are contained within $\eta^{ns_1s_2}_{jj'}$. For each order $n$ with harmonics $s_1$ and $s_2$ satisfying $n\ge|s_1|+|s_2|$, we are thus dealing with a self-consistent system in $\phi^{ns_1s_2}$, which is treated using the approach described in Ref.\ \cite{paper247}. 

Specifically, we proceed by first using the identity $\rho_{ll'}^0=\sum_{jj'}a_{jl}a_{j'l'}\tilde{\rho}_{jj'}^0=\sum_ja_{jl}a_{jl'}f_j$ in the sum of Eq. (\ref{rhons2}), and then moving from state to site representation to obtain the diagonal density-matrix elements as
\begin{align}\label{rhonsll2}
\rho^{ns_1s_2}_{ll}=&\frac{-1}{2e}\sum_{l'}\chi^{0}_{ll'}\left(s_1\omega_1+s_2\omega_2\right)\phi^{ns_1s_2}_{l'} \\
&+\sum_{jj'}a_{jl}a_{j'l}\eta^{ns_1s_2}_{jj'}, \nonumber
\end{align}
where
\begin{equation}\label{chi0}
\chi^{0}_{ll'}\left(\omega\right)=\frac{2e^2}{\hbar}\sum_{jj'}\left(f_{j'}-f_j\right)\frac{a_{jl}a_{j'l}a_{jl'}a_{j'l'}}{\omega+\ii/2\tau-\left(\varepsilon_j-\varepsilon_{j'}\right)}
\end{equation} 
is the noninteracting RPA susceptibility at frequency $\omega$.

In summary, each new iteration order $n$ is computed from the results of previous orders following a precedure similar to Ref.\ \cite{paper247}, here generalized to deal with multifrequency cw illumination:
\begin{enumerate}
\item We first calculate $\eta^{ns_1s_2}_{jj'}$ using Eq.\ (\ref{etajj}).
\item We then combine Eqs.\ (\ref{phinsrho}) and (\ref{rhonsll2}) to find a self-consistent equation for $\phi_l^{ns_1s_2}$, which reduces in matrix form to
\begin{equation}\nonumber
\phi^{ns_1s_2}=\left[1-v\cdot\chi^{0}\left(s_1\omega_1+s_2\omega_2\right)\right]^{-1}\cdot\beta^{ns_1s_2},
\end{equation}
using site labels $l$ as matrix indices and having defined
\begin{align}\nonumber
\beta^{ns_1s_2}_l=&-(\textbf{R}_l\cdot\eh)\;E_0\delta_{n,1}\sum_{i=1}^{2}(\delta_{s_i,-1}+\delta_{s_i,1}) \\
&-2e\sum_{l'jj'}v_{ll'}a_{jl}a_{j'l}\eta_{jj'}^{ns_1s_2}. \nonumber
\end{align}
\item We use the calculated values of $\eta_{jj'}^{ns_1s_2}$ and $\phi^{ns_1s_2}_l$ to obtain $\rho^{ns_1s_2}_{ll}$ using Eq.\ (\ref{rhonsll2}), and from here the induced charge at site $l$ at order $n$ associated with the harmonics $s_1$ and $s_2$ as $\rho^{\rm ind}_l=-2e\rho^{ns_1s_2}_{ll}$. Computation and storage demand are reduced by using the property $\tilde{\rho}_{jj'}^{ns_1s_2}=\left(\tilde{\rho}_{j'j}^{n,-s_1,-s_2}\right)^*$.
\item Finally, the polarizability for wave mixing among the harmonics $s_1$ and $s_2$ is calculated from
\begin{equation}
\alpha^{(n)}\left(s_1\omega_1+s_2\omega_2\right)=-\frac{2e}{(E_0)^n}\sum_l\rho^{ns_1s_2}_{ll}\;\Rb_l \cdot \eh
\end{equation}
upon iteration of this procedure up to order $n$.
\end{enumerate}
Extension to more frequencies is straightforward.

\section{Effect of doping on wave mixing}

\begin{figure*}
\includegraphics[width=0.9\textwidth]{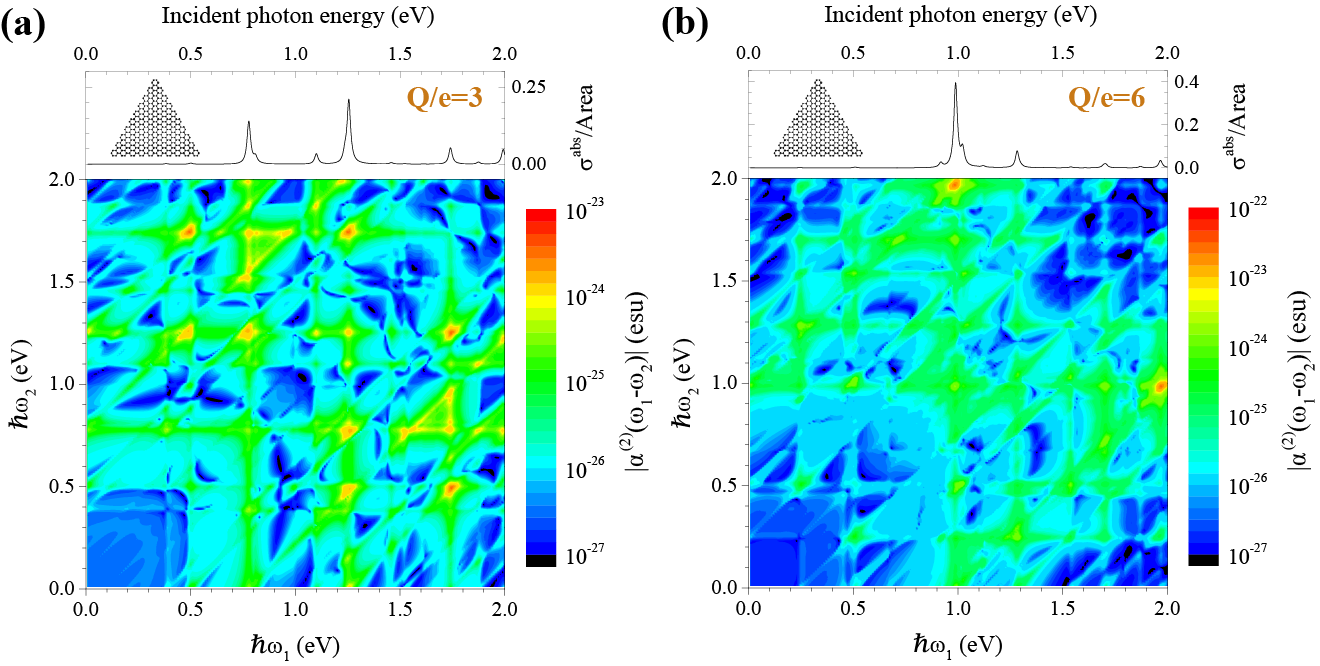}
\caption{\label{figS3} \textbf{Difference frequency generation.} The nonlinear polarizability $\alpha^{(2)}(\omega_1-\omega_2)$, corresponding to difference frequency generation, is shown for a triangular, armchair-edged graphene nanoisland containing $N=330$ carbon atoms and doped with (a) three and (b) six additional electrons. In each case the upper panels in (a) and (b) show the linear absorption cross-section of the nanoisland, while the lower panels show the polarizabilities as the incident field frequencies $\omega_1$ and $\omega_2$ are varied.}
\end{figure*}

\begin{figure*}
\includegraphics[width=0.9\textwidth]{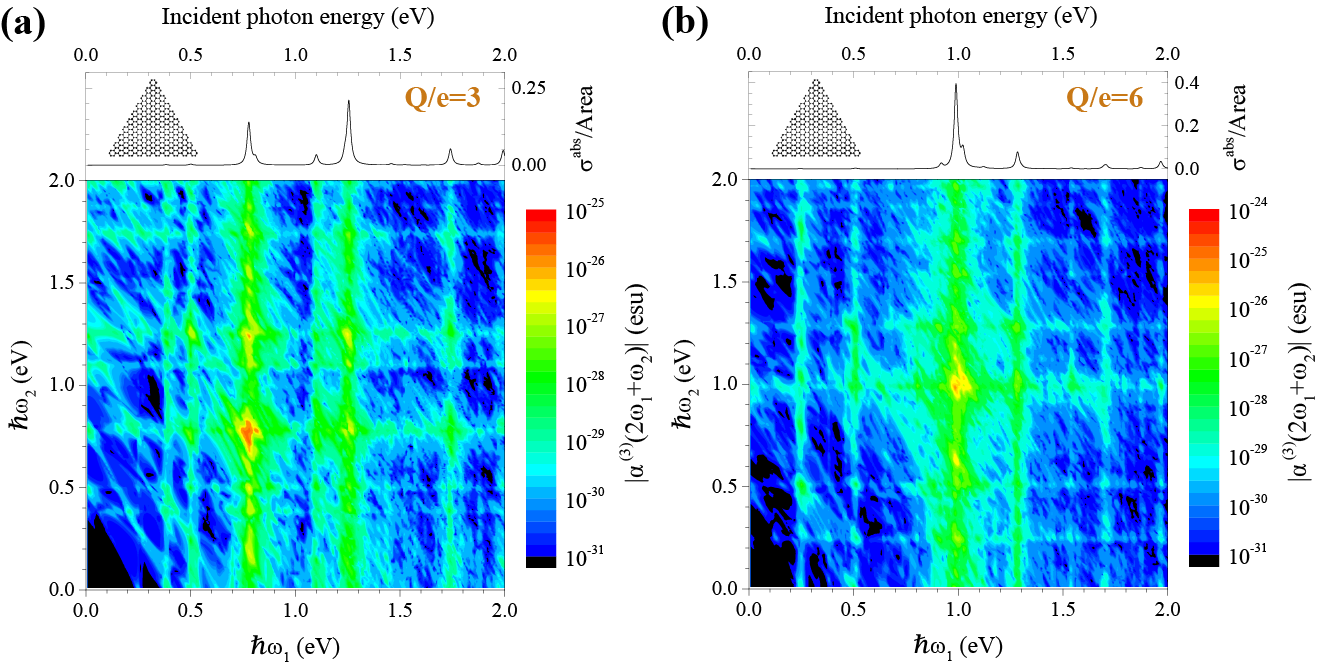}
\caption{\label{figS4} \textbf{Four-wave mixing.} The nonlinear polarizability $\alpha^{(3)}(2\omega_1+\omega_2)$, corresponding to four-wave mixing, is shown for the same nanoisland and doping levels considered in Fig.\ \ref{figS3}.}
\end{figure*}

The linear absorption spectrum of an armchair-edged, doped graphene nanoisland exhibits numerous plasmonic resonance features, some of which display strong electrical tunability. In particular, the lowest-energy prominent peak, which is absent when the nanoisland is undoped, undergoes dramatic energy shifts upon adding only a few additional charge carriers (see Fig.\ \ref{figS1}). As this mode is tunable, and also tends to produce large nonlinearities \cite{paper247}, we excite it in the nanoisland considered in Fig.\ 1 of the main text with one of two collinear ultrashort pulses to study wave mixing. In Fig.\ \ref{figS2} we investigate wave mixing of collinear pulses in the same nanoisland, but here one of the pulses is tuned to a different plasmon from that of the main text. Interestingly, a different qualitative behavior is observed when the island is either undoped (Fig.\ \ref{figS2}a) or doped (Fig.\ \ref{figS2}b). The former indicates that the second-order response vanishes for polarizations both parallel (upper panel) and perpendicular (lower panel) to an edge of the nanoisland when it is undoped, while a strong third-order response remains, although it is weaker than that of Fig.\ 1b. When the nanoisland is doped, the second-order response is recovered for the perpendicular polarization, but overall the nonlinear response is weaker than that of Fig.\ 1b, where the highly-tunable mode participates in wave mixing.

\section{Additional wave mixing polarizabilities}

In Figs.\ \ref{figS3} and \ref{figS4} we present the wave-mixing polarizabilities corresponding to difference frequency generation, $\alpha(\omega_1-\omega_2)$ and four-wave mixing, $\alpha(2\omega_1+\omega_2)$, respectively, for the $N=330$ graphene nanoisland considered in Figs.\ 2 and 3 of the main text doped with (a) three or (b) six electrons. As before, we find similar plasmonic enhancement of the input frequencies in both cases (see horizontal and vertical features in the contour plots), while in Fig.\ \ref{figS3} enhancement at the output frequency follows the curve $\omega_2=\omega_1+\omega_p$ and in Fig.\ \ref{figS4} the output enhancement follows $\omega_2=-2\omega_1+\omega_p$, $\omega_p$ being any of the plasmonic resonances in an island. In Fig.\ \ref{figS3} we note a very large difference frequency generation polarizability near $\hbar\omega_1\approx1\,$eV, $\hbar\omega_2\approx2\,$eV (or vice versa), where in fact we have a triple-resonance condition: simultaneously, both of the input frequencies and the output frequency are resonant with plasmons in the nanoisland. In Fig.\ \ref{figS4}, the cases where $\omega_1=\omega_2=\omega_p$ correspond to plasmonic enhancement of third-harmonic generation (THG). For comparison, the third-harmonic susceptibility $\chi^{(3)}(3\omega)$ has been measured in 10\,nm silver nanoparticles as $\sim2\times10^{-11}\,$esu \cite{LTY06}, corresponding to a third-harmonic polarizability $\alpha^{(3)}(3\omega)$ of $\sim2\times10^{-30}\,$esu.

\providecommand*{\mcitethebibliography}{\thebibliography}
\csname @ifundefined\endcsname{endmcitethebibliography}
{\let\endmcitethebibliography\endthebibliography}{}

\end{document}